\documentclass[a4paper]{article}

\usepackage{INTERSPEECH2019}
\usepackage{epstopdf}
\usepackage{epsfig}

\title{VAE-based regularization for deep speaker embedding}
\name{Yang Zhang$^1$$^,$$^2$, Lantian Li$^1$, Dong Wang$^{1*}$}
\address{
  $^1$Center for Speech and Language Technologies, Tsinghua University, China\\
  $^2$Beijing University of Posts and Telecommunications, China}
\email{\{zhangyang,lilt\}@cslt.org; wangdong99@mails.tsinghua.edu.cn}

\begin{document}

\maketitle
\begin{abstract}

Deep speaker embedding has achieved state-of-the-art performance in speaker recognition.
A potential problem of these embedded vectors (called `x-vectors') are not Gaussian,
causing performance degradation with the famous PLDA back-end scoring.
In this paper, we propose a regularization approach based on Variational Auto-Encoder (VAE).
This model transforms x-vectors to a latent space where mapped latent codes are more Gaussian,
hence more suitable for PLDA scoring.

\end{abstract}

\section{Introduction}

Automatic speaker verification (ASV) has found a broad range of applications.
Conventional ASV methods are based on statistical models~\cite{Reynolds00,Kenny07,dehak2011front}.
Perhaps the most famous statistical model in ASV is the Gaussian mixture model-universal background
model (GMM-UBM)~\cite{Reynolds00}.
This model represents the `main' variance of speech signals by a set of global Gaussian components (UBM), and the speaker characters are
represented as the `shift' of speaker-dependent GMMs over each Gaussian component of the UBM, denoted by
a `speaker supervector'.
The GMM-UBM architecture was later enhanced by subspace models, which assume that a speaker supervector can be factorized
into a speaker vector (usually low-dimensional) and a residual that represents intra-speaker variation.
The joint factor analysis~\cite{Kenny07,kenny2005joint} was the most successful subspace model in early days,
though the following i-vector model obtained more attention~\cite{dehak2011front}. Besides the simple
structure and the superior performance, the i-vector approach firstly demonstrated that a speaker can be
represented by a low-dimensional vector, which is the precursor of the important concept of \textbf{speaker embedding}.

It should be emphasized, however, that the i-vector model is purely unsupervised and the embeddings (i-vectors)
contain a multitude of variations more than speaker information. Therefore, it heavily relies on a powerful back-end scoring model
to achieve reasonable performance. Among various back-end models, the PLDA model~\cite{Ioffe06,prince2007probabilistic} has been
very powerful, in particular with a simple whitening and length normalization~\cite{garcia2011analysis}.
In the nut shell, PLDA assumes the `true' speaker codes within an i-vector is low dimensional and follows
a simple Gaussian prior, and the residual is a full-rank Gaussian, formally written by:

\begin{equation}
\label{eq:plda}
\phi_{su} = m + {\rm U}y_s + \epsilon_{su},
\end{equation}

\noindent where $\phi_{su}$ is the i-vector of utterance $u$ of speaker $s$, $y_s \sim N(0, I)$  and $\epsilon_{su} \sim N(0, W)$
are speaker codes and residual respectively, $m$ is the global shift and ${\rm U}$ is the speaker loading matrix. Under this assumption, the speaker prior $p(y_s)$, the conditional $p(\phi|y_s)$
and the marginal $p(\phi)$ are all Gaussian. Fortunately, i-vectors match these conditions pretty well, due to the linear Gaussian structure of the i-vector model.
Partly for this reason, the i-vector/PLDA framework remains a strong baseline on many ASV tasks.

Recently, neural-based ASV models have shown great potential~\cite{ehsan14,heigold2016end,li2017deep,snyder2018xvector}.
These models utilize the power of deep neural networks (DNNs) to learn strong speaker-dependent features, ideally from a large amount of speaker-labelled data. The present research can be
categorized into frame-based learning~\cite{ehsan14,li2017deep} and utterance-based learning~\cite{heigold2016end,snyder2018xvector,zhang2016end,snyder2016deep}.
The frame-based learning intends to learn short-time speaker features, thus more generally useful for
speaker-related tasks, while the utterance-based learning focuses on a whole-utterance speaker representation and/or classification,
hence more suitable for the ASV task. A popular utterance-based learning approach is the x-vector model proposed by Snyder et al.~\cite{snyder2018xvector},
where the first- and second-order statistics of frame-level features are collected and projected to a low-dimensional
representation called x-vector, with the objective of discriminating the speakers in the training dataset.
The x-vector model has achieved good performance in various speaker recognition tasks, as well as related tasks such as
language identification~\cite{Snyder2018}. Essentially, the x-vector model can be regarded as a deep and discriminative
counterpart of the i-vector model, and is often called \textbf{deep speaker embedding}.


Interestingly, experiments show that the x-vector system also heavily relies on a strong back-end scoring model, in particular
PLDA. Since the x-vector have been sufficiently discriminative, the role of PLDA here is \textbf{regularization} rather
than \textbf{discrimination} (as in the i-vector paradigm): it (globally) discovers the underlying speech codes that are intrinsically Gaussian,
so that the ASV scoring based on these codes tends to be comparable across speakers. A potential problem, however,
is that x-vectors inferred from DNNs are unconstrained, which means that the speaker distribution and the speaker conditional
could be in any form. These unconstrained distributions may cause great difficulty for PLDA to
discover the underlying speaker codes that are assumed to be Gaussian. Some researchers have noticed this problem and proposed some remedies that encourage speaker conditionals more Gaussian~\cite{Cai2018,li2019gaussian}, but none of them
constrain the prior, thus produced x-vectors are still not suitable for PLDA modeling.

In this paper, we investigate an explicit regularization model for unconstrained x-vectors. This model is inspired by the
variational auto-encoder (VAE) architecture, which is capable of projecting an unconstrained distribution to a simple
Gaussian distribution. This can be used to constrain the marginal distribution of x-vectors. Moreover, a
cohesive loss is added to the VAE objective. This follows the same spirit of~\cite{Cai2018,li2019gaussian} and can
constrain the speaker conditionals. Experiments showed that with this VAE-based regularization, performance of cosine scoring
is largely improved, even comparable with PLDA. This indicates that VAE plays a similar role as PLDA, or,
in other words, PLDA works as a regularizer rather than a discriminator in the x-vector scoring.
Furthermore, the VAE-based speaker codes achieved the state-of-the-art performance when scoring with PLDA,
demonstrating that (1) VAE-based speaker codes are more regularized and suitable for PLDA modeling, and (2) VAE-based regularization and PLDA scoring are complementary.

The organization of this paper is as follows. Section~\ref{sec:method} presents the VAE-based regularization model, and
the experiments are reported in Section~\ref{sec:exp}.
The paper is concluded in Section~\ref{sec:con}.

\section{VAE-based speaker regularization}
\label{sec:method}

\subsection{Revisit PLDA}

The principle of PLDA is to model the marginal distribution of speaker embeddings (i-vector or x-vector), by
factoring the total variation of the embeddings into between-speaker variation and within-speaker
variation. Based on this factorization, the ASV decision can be cast to a hypothesis test~\cite{Ioffe06,prince2007probabilistic},
formulated by:

\begin{eqnarray}
s(\phi_1, \phi_2) &=& \frac{P(\phi_1 = \phi_2 | \Lambda )}{P(\phi_1 \neq \phi_2 | \Lambda )} \nonumber \\
                  &=& \frac{\int p(\phi_1,\phi_2|y) p(y) {\rm d}y}{\int p(\phi_1|y) p(y) {\rm d}y \int p(\phi_2|y) p(y) {\rm d}y}, \nonumber
\end{eqnarray}

\noindent where $s$ denotes the confidence score, and the equality relation ($\phi_1=\phi_2$) denotes that the two embeddings are from the same
speaker.

According to Eq.(\ref{eq:plda}), PLDA is a linear Gaussian model and the prior, the conditional, and the marginal are Gaussian.
If the embeddings do not satisfy this condition, PLDA cannot model them well,
leading to inferior performance. This is the case of x-vectors, which are derived from DNNs and both the speaker prior and speaker
conditionals are unconstrained.
In order to deal with the unconstrained distributions of x-vectors, we need
a probabilistic model more complex than PLDA.

\subsection{VAE for regularization}

\begin{figure}[htbp]
\centering\includegraphics[width=0.7\linewidth]{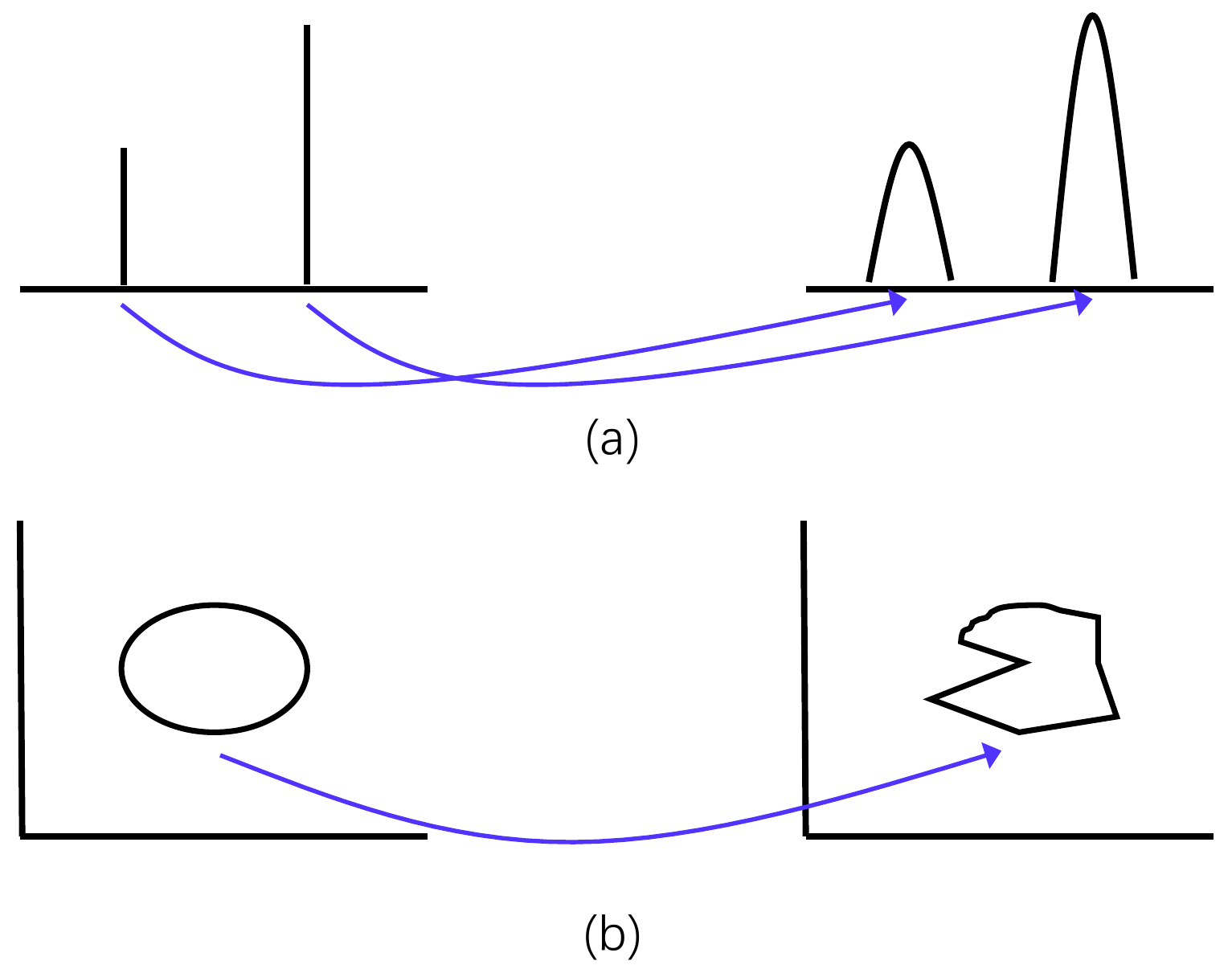}
\caption{Two examples of distribution mapping. (a) a discrete distribution is mapped to a mixture of two Gaussians; (b) a 2-dim Gaussian is mapped to an irregular distribution.}
\label{fig:map}
\end{figure}

VAE is a generative model (like PLDA) that can represent a complex data distribution~\cite{kingma2013auto}. The key idea of VAE is to learn a DNN-based mapping function
$x=f(z)$ that maps a simple distribution $p(z)$ to a complex distribution $p(x)$. In other words, it represents
 complex observations by simple-distributed latent codes via \textbf{distribution mapping}. An illustration of this mapping is shown in Fig.~\ref{fig:map}.
It can be easily shown that the mapped distribution is written by:

\[
\log p(x)  = \log p(z)  + \log |{\rm det} \frac{{\rm d}f^{-1}(x)}{{\rm d}x}|,
\]

\noindent where $f^{-1}$ is the inverse function of $f(z)$.

Although VAE can be used to represent the complex \textbf{marginals}, it does not involve any class structure, and so cannot be
used in the hypothesis test scoring framework. Nevertheless, if we can find the posterior $p(z|x)$, the complex $p(x)$ can be
mapped to a more constrained $p(z)$, so the simple cosine distance can be used for verification. Moreover, the regularized
code $z$ tends to be easily modeled by PLDA, hence combining the strength of VAE in distribution mapping and the strength of PLDA in
distinguishing between- and within-speaker variations. Fortunately, VAE provides a simple way to infer an approximation distribution of $p(z|x)$, denoted by $q(z|x)$.
It learns a function $g(x)$, parameterized by a DNN, to map $x$ to the parameters of $q(z|x)$, which are the mean and covariance if $q(z|x)$ is assumed to be Gaussian.
By this setting, the mean vector of $q(z|x)$ can be treated as VAE-regularized speaker codes, and can be used in cosine or PLDA-based scoring.

\begin{figure}[htbp]
\centering\includegraphics[width=\linewidth]{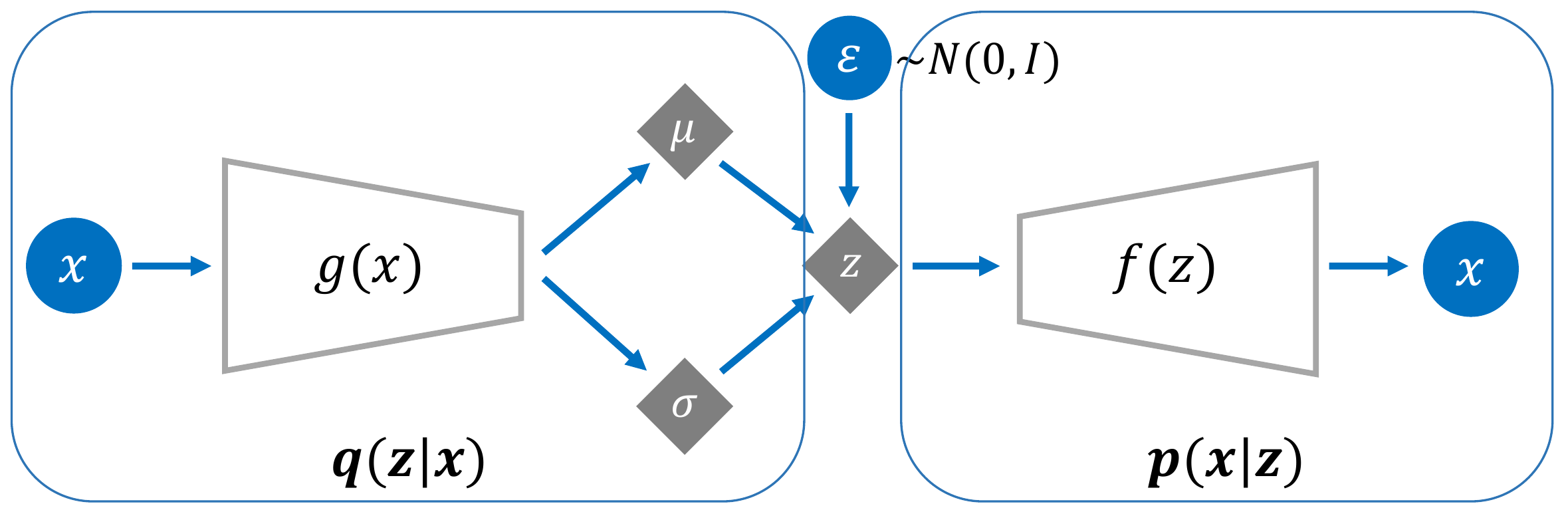}
\caption{The standard VAE architecture. It involves a generative model (decoder) $p(z)$ and $p(x|z)$, and an inference model (encoder) $q(z|x)$ that approximates the posterior $p(z|x)$. Both $p(x|z)$ and
$q(z|x)$ involve a DNN-based mapping function. A random variable $\epsilon$ is used to facilitate the construction of $q(z|x)$, known as `reparameterization trick'~\cite{kingma2013auto}.}
\label{fig:vae1}
\end{figure}

Fig.~\ref{fig:vae1} illustrates the VAE framework. In this framework, a \textbf{decoder} $f(z)$ maps $p(z)$ to $p(x)$, i.e.,

\[
p(x) = \int p(x|z) p(z) {\rm d}z  = \int N(f(z), I) p(z) {\rm d}z,
\]

\noindent where $p(x|z)$ has been assumed to be a Gaussian. Furthermore, an \textbf{encoder} $g(x)$ produces a distribution $q(z|x)$
that approximates the posterior distribution $p(z|x)$ as follows:

\[
p(z|x) \approx q(z|x) = N(\mu(x),\sigma(x)),
\]

\noindent where $[\mu(x)\ \ \sigma(x)] = g(x)$.

The training objective is the log probability of the training data $\sum_i \ln p(x_i)$.
It is intractable so a variational lower bound is optimized instead, which depends on both the encoder $g(x)$ and the decoder $f(z)$. This is formally written by:

\[
L(f,g) = \sum_i \{ -D_{\rm KL} [q(z|x_i) || p(z)] + \mathbb{E}_{q(z|x_i)} [\ln p(x_i|z)] \},
\]

\noindent where $D_{\rm KL}$ is the KL distance, and $\mathbb{E}_{q}$ denotes expectation w.r.t. distribution $q$. As
the expectation is intractable, a sampling scheme is often used, as shown in Fig.~\ref{fig:vae1}.
More details of the training process can be found in~\cite{kingma2013auto}.

Note that the $L(f,g)$ involves two components: a regularization term that pushes $q(z|x)$ to $p(z)$, and
a reconstruction term that encourages a good reconstruction of $x$ from $z$. We are free to tune the relative weights
of these two terms in practice, in order to obtain latent codes that are either more regularized or more representative.
This freely-modified objective may be never a variational lower bound, though non-balanced weights often lead to
better performance in our experiments.

\subsection{Speaker cohesive VAE}

\begin{figure}[htbp]
\centering\includegraphics[width=\linewidth,height=3.7cm]{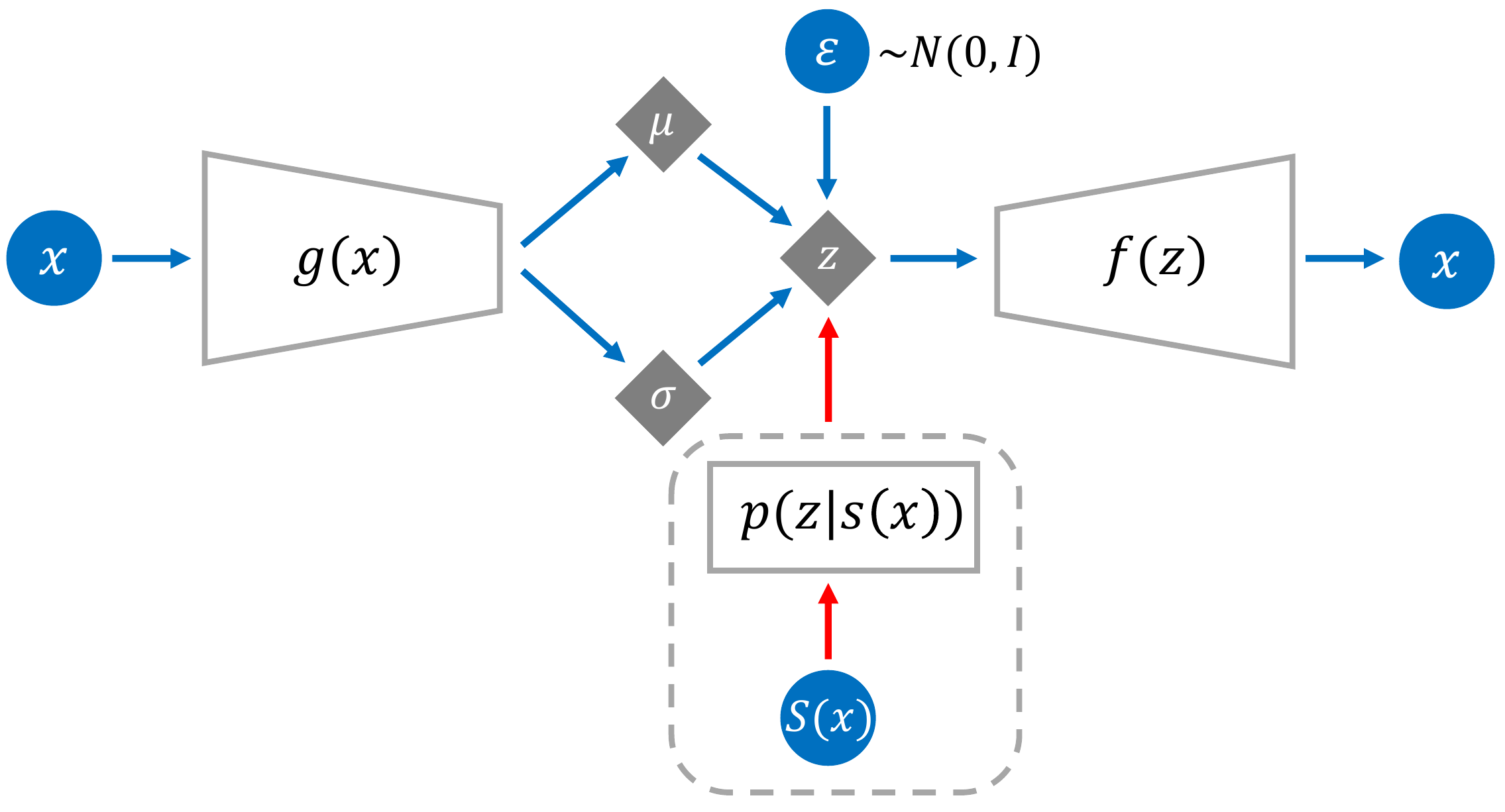}
\caption{The VAE architecture with cohesive loss (dotted box). }
\label{fig:vae2}
\end{figure}

The standard VAE only constrains the marginal distribution $p(z)$ to be Gaussian, which does not guarantee a Gaussian prior or a Gaussian conditional.
This is because the VAE model is purely unsupervised and there is no speaker information involved. This lack of speaker information is probably not
a big issue for x-vectors as they are speaker discriminative already. However, considering speaker information may help VAE
to produce a better regularization. Especially, if the speaker code $z_{s,u}$ of a particular speaker $s$ can be regularized to be Gaussian, the scores
based on either cosine distance or PLDA will be more across-speaker comparable. This can be formulated as an additional term in the VAE objective
function, which we call \textbf{speaker cohesive loss} denoted by $L_{\rm C}$:

\[
L_{\rm C}(f,g) = \sum_i \ln p(\mu(x)|s(x)) = \sum_i \ln N(\mu(x); s(x), I)
\]
\noindent where $s(x)$ denotes the mean of $\mu(x)$ of all utterances that belong to the same speaker as $x$. This essentially follows the
same spirit of the central loss used in~\cite{Cai2018,li2019gaussian}. Fig.~\ref{fig:vae2} illustrates the VAE architecture
with cohesive loss, and we name this improved VAE architecture as \textbf{Cohesive VAE}.

\section{Experiments}
\label{sec:exp}

\subsection{Data}

Three datasets were used in our experiments: VoxCeleb, SITW and CSLT-SITW.
VoxCeleb was used for model training, while the other two were used for evaluation.
More information about these three datasets is presented below.

\emph{VoxCeleb}: A large-scale free speaker database collected by University of Oxford, UK~\cite{nagrani2017voxceleb}.
The entire database involves \emph{VoxCeleb1} and \emph{VoxCeleb2}.
This dataset, after removing the utterances shared by SITW, was used to train the x-vector model, plus the PLDA and VAE models.
Data augmentation was applied, where the MUSAN corpus~\cite{musan2015} was used to generate noisy utterances and
the room impulse responses (RIRS) corpus~\cite{ko2017study} was used to generate reverberant utterances.

\emph{SITW}: A standard database used to test ASV performance in real-world conditions~\cite{mclaren2016speakers}.
It was collected from open-source media channels, and consists of speech data covering $299$ well-known persons.
There are two standard datasets for testing: \emph{Dev. Core} and \emph{Eval. Core}. We used \emph{Dev. Core} to select model parameters,
and \emph{Eval. Core} to perform test in our first experiment.
Note that the acoustic condition of SITW is similar to that of the training set VoxCeleb, so this test can
be regarded as an \textbf{in-domain test}.

\emph{CSLT-SITW}: A small dataset collected by CSLT for commercial usage. It consists of $77$ speakers,
each of which records a simple Chinese command word, and the duration is about $2$ seconds.
The scenarios involve laboratory, corridor, street, restaurant, bus, subway, mall, home, etc.
Speakers varied their poses during the recording, and the recording devices were placed both near and far.
There are about $30k$ utterances in total. The acoustic condition of this dataset is quite different from that of the training set VoxCeleb, and was used for \textbf{out-of-domain test}.

\subsection{Settings}

We built several systems to validate the VAE-based regularization, each involving a particular pair of front-end and back-end.

\subsubsection{Front-end}

\textbf{x-vector}: The baseline x-vector front-end. It was built following the Kaldi SITW recipe~\cite{povey2011kaldi}.
The feature-learning component is a $5$-layer time-delay neural network (TDNN).
The statistic pooling layer computes the mean and standard deviation of the frame-level features from a speech segment.
The size of the output layer is $7,185$, corresponding to the number of speakers in the training set.
Once trained, the $512$-dimensional activations of the penultimate hidden layer are read out as an x-vector.
\\
\textbf{v-vector}: The VAE-regularized speaker code. The VAE model is a $7$-layer DNN. The dimension of code layer is $200$, and other hidden layers are $1,800$.
The x-vectors of all the training utterances are used to the VAE training.
\\
\textbf{c-vector}: The VAE-regularized speaker code, with the cohesive loss involved in the VAE training. The model structure is the same as
in the v-vector front-end, and the v-vector VAE was used as the initial model for training.
We tuned the weight $L_{\rm C}(f,g)$ in the objective function, and found $10$ is a reasonable value.
\\
\textbf{a-vector}: Speaker code regularized by a standard auto-encoder (AE). AE shares a similar
structure as VAE, but the latent codes are not probabilistic so it is less capable of modeling complex distributions.
The AE structure is identical to the VAE model in the v-vector front-end, except that the
code layer is deterministic.

\subsubsection{Back-end}

\textbf{Cosine}: Simple cosine distance.
\\
\textbf{PCA}: PCA-based projection (150-dim) plus cosine distance.
\\
\textbf{PLDA}: PLDA scoring.
\\
\textbf{L-PLDA}: LDA-based projection (150-dim) plus PLDA scoring.
\\
\textbf{P-PLDA}: PCA-based projection (150-dim) plus PLDA scoring.

\subsection{In-domain test}

The results on the two SITW evaluation sets, \emph{Dev. Core} and \emph{Eval. Core}, are reported
in Table~\ref{tab:sitw}. The results are reported in terms of
equal error rate (EER).

Firstly focus on the x-vector front-end. It can be found that PLDA scoring outperformed cosine distance.
As we argued, this cannot be interpreted as the discriminative nature of PLDA, but its regularization
capability.
This is supported by the observation that the v-vector front-end
achieved rather good performance with the cosine back-end (compared with x-vector + PLDA).
Since VAE is purely unsupervised, it only contributes to regularization. This suggests that PLDA may play a
similar role as VAE.

Secondly, we observe that with PCA or LDA, PLDA can perform much better. It is not convincing to assume that
LDA and PCA improve the discriminant power of x-vectors (in particular PCA), so the only interpretation is that
these two models performed regularization, generating more Gaussian codes that are suitable for PLDA. This
regularization is similar as what VAE did, but it seems VAE did a better job than PCA, and even better than LDA on the
larger evaluation set \emph{Eval. Core}, even without any speaker supervision.

Thirdly, it can be found that c-vectors performed better than v-vectors with cosine scoring, confirming that
involving cohesive loss improves the regularization. When combined with PLDA, however, the advantage of c-vectors
diminished. This is expected as PLDA has already learned the speaker discriminative knowledge.

Finally, we found that other unsupervised regularization methods, including PCA and AE, can not obtain reasonable
performance with cosine distance, indicating that they cannot conduct good regularization by themselves. This
is contrast to VAE, confirming the importance of the probabilistic codes: without this probabilistic nature, it would
be impossible to model the complex distribution of x-vectors.

\begin{table}[htb!]
 \begin{center}
  \caption{Performance (EER\%) on SITW Dev. Core and Eval. Core.}
  \vspace{-2mm}
   \label{tab:sitw}
     \begin{tabular}{|l|c|c|c|c|c|}
                                 \multicolumn{6}{c}{SITW Dev. Core} \\
      \hline
                                 &  Cosine   &   PCA   &   PLDA     &   L-PLDA    &   P-PLDA   \\
       \hline
             x-vector            &  15.67    &  16.17  &   9.09     & \textbf{3.12} &   4.16    \\
       \hline
             a-vector            &  16.10    &  16.48  &   11.21    &   4.24      &   5.01    \\
       \hline
             v-vector            &  10.32    &  9.94   &   3.62     &   3.54      &   4.31    \\
       \hline
             c-vector            &\textbf{9.05}&\textbf{8.55}&\textbf{3.50}& 3.31 &\textbf{3.85} \\
       \hline

                                       \multicolumn{6}{c}{SITW Eval. Core} \\
       \hline
                                 &  Cosine   &   PCA   &   PLDA     &   L-PLDA      &   P-PLDA   \\
       \hline
             x-vector            &  16.79    &  17.22  &   9.16     &   3.80        &   4.84   \\
       \hline
             a-vector            &  16.05    &  16.81  &   12.14    &   4.27        &   5.09    \\
       \hline
             v-vector            &  10.11    &  10.03  &\textbf{3.64}&   3.64        &   4.43    \\
       \hline
             c-vector            &\textbf{9.05}&\textbf{8.83}&   3.77 & \textbf{3.53}& \textbf{4.10} \\
       \hline
     \end{tabular}
 \end{center}
\end{table}

\vspace{-3mm}

\subsection{Analysis}

To better understand the VAE-based regularization, we compute the skewness and kurtosis of the distributions
of different speaker codes. The skewness and kurtosis are defined as follows:

\[
{\rm Skew}(x) = \frac {E[(x-\mu_x)^3]}{\sigma_x^3} \ , \ {\rm Kurt}(x)=\frac{E[x-\mu_x]^4}{\sigma_x^4} - 3,
\]

\noindent where $\mu_x$ and $\sigma_x$ denote the mean and standard variation of $x$, respectively.
More Gaussian is a distribution, more close to zero are the two values.

The utterance-level and speaker-level skewness and kurtosis of different speaker codes are reported in Table~\ref{tab:gauss}.
Focusing on the utterance-level results, it can be seen that the values of skewness and kurtosis of both v-vector and c-vector are clearly
smaller than x-vector. This means that the v-vector and the c-vector are more Gaussian.
For the speaker-level results, it can be found that the kurtosis was largely reduced in v-vectors and c-vectors.
This indicates that the Gaussian regularization placed by VAE on the marginal has implicitly regularized the prior, which
is the major reason that these vectors are more suitable for PLDA.
The a-vector, derived from AE, has smaller skewness but larger kurtosis compared to the x-vector, on both the utterance-level and the
speaker-level, suggesting that AE did not perform a good regularization.

\begin{table}[htb!]
 \begin{center}
  \caption{Utterance-level and speaker-level skewness and kurtosis of different speaker codes on the Voxceleb set.}
  \vspace{-2mm}
   \label{tab:gauss}
     \begin{tabular}{|l|c|c||c|c|}
      \hline
                                 &  Skew(utt)   &   Kurt(utt)   &  Skew(spk)   &   Kurt(spk)  \\
       \hline
             x-vector            &  -0.0423     &   -0.3604     &  0.0018      &   -0.4499    \\
       \hline
             a-vector            &  -0.0072     &   -0.7740     &  0.0014      &   -0.9765     \\
       \hline
             v-vector            &  -0.0055     &   0.1324      &  -0.0042     &   -0.0285     \\
       \hline
             c-vector            &  -0.0043     &   0.1154      &  -0.0076     &   -0.0298    \\
       \hline
     \end{tabular}
 \end{center}
\end{table}

\vspace{-3mm}

\subsection{Out-of-domain test}

In this experiment, we test the performance of various systems on the CSLT-SITW dataset. Due to the
limited data, three-fold cross-validation was used whenever training is required.
Three experiments were conducted:
(1) directly using all the front-end and back-end models trained by VoxCeleb; (2) retraining all the models except the x-vector DNN;
(3) the same as the retraining scheme, but all the PLDA models were trained by an unsupervised adaptation~\cite{garcia2014improving}. The results
show that scheme (2) is generally the best, and the PLDA adaptation contributes additional gains in some test settings.
For simplicity, only the retraining results under scheme (2) are reported in Table~\ref{tab:cslt}.
The results exhibit a similar trend as in the SITW test, that both the v-vector and c-vector outperform the x-vector, and the c-vector obtained the best performance in nearly all the test settings. Compared to the SITW test, the larger performance
gains obtained by VAE-regularization. It might be attributed to the more complex acoustic conditions of CSLT-SITW, though more
investigation is required.

\begin{table}[htb]
 \begin{center}
  \caption{Performance (EER\%) on CSLT-SITW.}
    \vspace{-2mm}
   \label{tab:cslt}
   \scalebox{1}{
     \begin{tabular}{|l|c|c|c|c|c|c|}

       \hline
                               &   Cosine  &    PCA    &   PLDA    &  L-PLDA  &  P-PLDA   \\
       \hline
             x-vector          &   16.65   &   16.89   &   16.91   &   15.39  &   13.29   \\
       \hline
             v-vector          &   13.55   &   13.71   &\textbf{12.46}&   12.06  &   12.02  \\
       \hline
             c-vector          &\textbf{12.98}&\textbf{13.13}& 12.48 & \textbf{12.01}  & \textbf{11.98}  \\
       \hline
     \end{tabular}
   }
 \end{center}
\end{table}

\vspace{-3mm}





\section{Conclusions}
\label{sec:con}

This paper proposed a VAE-based regularization for deep speaker embedding. By this model, x-vectors that usually exhibit a complex distribution are mapped to
latent speaker codes that are simply Gaussian. This model
was further enhanced by a speaker cohesive loss, which regularizes speaker conditionals. Experiments on the SITW dataset and a private commercial dataset
demonstrated that
the VAE-regularized speaker codes can achieve better performance with either cosine distance or PLDA scoring, compared to the x-vector baseline.
Future work will investigate speaker-aware VAE, where speaker codes and utterance codes are
hierarchically linked as in PLDA.



\bibliographystyle{IEEEtran}

\bibliography{vae}

\end{document}